# Swimming Cylinder Wake Control with Plasma Actuator

Javad Omidi [1,*]

[1] Department of Chemical Engineering, Columbia University, New York, USA
* Correspondence: jo2668@columbia.edu

**Abstract:** In this research, the effect of utilizing a micro plasma actuator on controlling the flow through a two-dimensional cylinder is investigated using an advanced electrostatic model. Within this model, by solving two elliptic equations, the potential distribution and plasma distribution in the solution domain are determined, leading to the generation of a volumetric force. This force is added to the momentum equations as a source term. The Reynolds number of the flow is set at 20,000, and the plasma actuator operates in a steady manner. Due to the high Reynolds number, the flow is turbulent, and its time-dependent nature is modeled as unsteady. Plasma actuators are symmetrically mounted at 45 and 90 degrees with respect to the free stream, both above and below the cylinder surfaces. The influence of the actuator placement angle on the flow quality downstream of the cylinder is examined. The results indicate an enhancement of flow quality by 8% to 15% downstream of the cylinder. Moreover, improvements of 40% to 55% in the variation of the lift coefficient and 75% to 90% in the drag coefficient are reported. The findings reveal superior performance of the actuator positioned at 90 degrees compared to the 45-degree orientation. This can be attributed to the proximity of the 90-degree actuator position to the point of boundary layer separation initiation. Furthermore, using the actuator positioned at 90 degrees and applying three different voltages of 10, 13, and 16 kV, the impact of flow control on the first cylinder in a tandem arrangement on the downstream flow of the second cylinder is examined. The results demonstrate an enhancement in downstream flow quality of 20% to 26%.

**Keywords:** Micro Plasma actuator; Electrostatic model; Turbulent flow control; 2D cylinder

## 1. Introduction

Today, the advancements achieved in the field of flow control have provided the opportunity to significantly enhance the performance of various human-made devices [1]. One of the fundamental physical phenomena investigated by researchers is the control of flow around a two-dimensional cylinder. Extensive studies have been conducted in this regard [2], employing various methods to control the shedding of von Kármán vortices at different Reynolds numbers. Among these methods, dielectric barrier discharge (DBD) plasma actuators are recognized as one of the most reliable and effective techniques [3]. Although less than two decades have passed since the invention of these actuators, researchers have evaluated their capabilities in various laboratory and industrial domains [4].

DBDs are remarkably simple yet highly effective. Their lightweight nature, which is crucial for applications involving high gravitational acceleration on the body, easy installation, lack of requirement for pneumatic, hydraulic, or movable systems, and very low power consumption are notable features [5]. In addition to these mentioned attributes, the rapid performance of these actuators in flow control should be emphasized [6 – 12]. The cost-effectiveness of utilizing these actuators and the possibility of their numerical modeling on an engineering scale are among other advantageous aspects [4].

The flow passing over a cylinder has long been a captivating and foundational fluid dynamic phenomenon studied by researchers. Due to the inherently unsteady nature of the flow after the Reynolds number surpasses 47, controlling the von Kármán vortices in these regimes has significant applications. Comprehensive reviews of flow passing over a cylinder have been conducted by Williamson [2] and Zdravkovich [13]. Furthermore, the flow passing over paired cylinders has diverse industrial applications, including simulating heat exchangers, cooling systems, and multiphase flow mixers. Zdravkovich [14] has conducted a thorough investigation of flow passing over various arrangements of twin cylinders at different Reynolds numbers. In 2006, Thomas et al. [15] carried out an experimental study using plasma actuators to control the flow passing over single cylinders and twin cylinders, aiming to reduce the noise of aircraft landing systems. Similarly, Kim et al. [16], in 2007, conducted tests to delay flow separation from a half-cylinder surface under atmospheric pressure. In another experimental study in 2008, Thomas, et al. [17] utilized plasma actuators to control instabilities and von Kármán vortex shedding from a cylinder. This control was aimed at noise reduction, with a Reynolds number of 33,000. In 2009, Corke et al. [18] presented a time-dependent model for plasma actuator modeling. One of the physical phenomena they examined was the passage of unsteady time-variable flow over a cylinder at Reynolds 40,000. In 2012, Mirzaei et al [19] conducted an experimental study concerning the effect of various plasma actuator shapes and electrodes on the lift coefficient behind a two-dimensional cylinder. Bernard and Moreau [20], in 2013, investigated the influence of steady and unsteady plasma actuators on downstream vortices behind a cylinder under different flow regimes.

In recent years, there has been an increased focus on research concerning the utilization of plasma actuators for controlling the flow passing over cylinders. In studies conducted by Bhattacharya and Gregory [21 – 23], plasma actuators using spanwise-segmented dielectric-barrier discharge were attached to cylinders to influence wake dynamics. Different power levels and configurations led to contrasting effects. Lower power caused larger-scale spanwise structures within the forcing region to precede non-forced ones, with deviations from the centerline and streamwise vorticity. Vortex shedding reduction was significant with high-power forcing. Al-Sadawi and Chong [24] used plasma actuators to reduce noise in single and tandem cylinder configurations. Activation of upstream and downstream cylinders concurrently lowered both narrowband and broadband noise. Joshi and Bhattacharya [25] employed large-eddy simulation to study segmented plasma forcing on wake attributes, focusing on three-dimensionality and separation changes. Plasma actuators were effective in suppressing vortex-induced vibrations (VIV) in a circular cylinder scenario [26]. Foshat [27] used a plasma actuator to control flow structure around a ground-effect plate, reducing separation regions. Chen and Wen [28] investigated plasma actuators and vortex generators for flow separation and vortex-induced vibration control, observing reduced turbulence, drag, and lift oscillation. Zhu et al. [29] found that plasma actuators effectively suppressed vortex shedding in the wake of a square cylinder when optimally positioned. Wang et al. [30] studied DBD plasma actuators' effects on a circular cylinder, revealing differences in drag reduction mechanisms between linear and sawtooth actuators.

Many of the current research endeavors regarding plasma actuators are conducted in an experimental manner, requiring significant expenses and time to investigate the real-world model. This circumstance has prompted ongoing advancements in various numerical models aimed at better simulating the effects of these actuators. Various models have been employed for simulating plasma actuators. The model presented by Roth [31], electrostatic models introduced by Suzen and Huang [32], circuit element models proposed by Orlov et al. [33], linearized force models suggested by Jayaraman and Anderson [34], Hall potential flow model [35], and full solution of Maxwell's equations without neglecting magnetic terms [36] have all been proposed for modeling the volumetric force induced by the plasma actuator. In this study, an enhanced version of the electrostatic model proposed by the authors is utilized [37]. This model accurately approximates the impact of the plasma actuator on fluid flow and stands as one of the most physically representative simulation models for the effects of plasma actuators [38 and 39].

The objective of this research is to employ an enhanced electrostatic model proposed by authors [37 – 39] for simulating the effect of a DBD plasma actuator on controlling the shedding of von Kármán vortices. The independent calibration of model parameters from experimental test results is a valuable feature incorporated in the development of this model, which is employed for the first time in the present research work. Additionally, in this investigation, the influence of plasma jet injection location on the boundary layer and the effect of varying applied voltage to the plasma actuator on the downstream flow quality are among the other objectives of this study.

## 2. Problem Definition

In this study, a DBD plasma actuator is utilized for the purpose of controlling the flow passing over a two-dimensional cylinder. This is achieved using an enhanced electrostatic numerical model developed by authors [37]. In this model, the effect of the plasma actuator is represented as a volumetric force exerted on the surrounding airflow, generating a wall jet by accelerating air particles. This volumetric force is added to the momentum equations through source terms in both the x and y directions. To generate this volumetric force, two additional equations will be coupled to the fundamental flow equations. These equations are solved iteratively and independently of the flow equations at each stage, contributing the physically generated force to the momentum equations. One of these equations represents the distribution of electric potential, while the other describes the distribution of charge density (produced plasma). Notable advantages of this model include the ease of modeling and the separation of plasma modeling equations from flow modeling equations, demonstrating sufficient accuracy in simulating the effect of the plasma actuator [38 and 39].

The airflow passing over a 1m diameter cylinder at a Reynolds number of 20,000 is simulated both with and without the use of a plasma actuator. An applied voltage of 10 kV at a frequency of 6 Hz is exerted on the electrodes. Plasma actuators are symmetrically mounted with respect to the flow direction at two angles: 45 and 90 degrees, positioned both above and below the cylinder. The effects of using the plasma actuator on enhancing flow quality downstream of the cylinder and the influence of actuator placement are studied. Figure 1 illustrates the arrangement of the plasma actuator on the cylinder at angles of 45 and 90 degrees.

In the second part of this simulation, the effect of the plasma actuator in its optimal placement on the upstream cylinder is analyzed for its impact on the downstream flow quality of a second cylinder in a tandem arrangement. In this study, two cylinders with equal diameters of 1m are positioned center-to-center at a separation of 4 cylinder diameters. The free-stream Reynolds number for this model is set at 20,000. In this investigation, the influence of increasing the voltage on the downstream flow quality is examined.

## 3. Computational Method

### 3.1. Fluid Dynamics Equations

Two-dimensional URANS equations have been employed to simulate the incompressible fluid flow. Since the primary energy input to the plasma actuator is directed towards accelerating fluid particles, and only a negligible fraction of it contributes to heating the fluid flow [40], the energy equation has been disregarded. The fundamental equations for mass conservation and momentum are the main focus. The governing equations for solving the fluid flow are as follows:

$$\frac{\partial u_j}{\partial x_j} = 0 \qquad (1)$$

$$\rho \frac{\partial u_i}{\partial t} + \rho u_j \frac{\partial u_i}{\partial x_j} = -\frac{\partial p}{\partial x_i} + v \frac{\partial^2 u_i}{\partial x_j \partial x_j} + f_{b_i} \qquad (2)$$

where, p is the static pressure, $\mu$ is the fluid viscosity, $u_j$ represents the velocity components, $\rho$ is the fluid density, $\vec{f_b}$ denotes the body force due to the plasma actuator per unit volume. As observed in Equation (2), the volumetric force resulting from the plasma actuator is added to the momentum equation as a source term.

### 3.2. Electrostatics Equations

When a high voltage is applied between two electrodes separated by a dielectric, a non-steady ionization process occurs. This non-steady ionization process takes place on nanosecond time scales and is continuously washed by the airflow [40], influencing the airflow on millisecond time scales [5]. The presence of charged particles resulting from these processes in the dominant electric field exerts a volumetric force on the external flow over the temporal scales of flow influence. The Lorentz volumetric force, while disregarding the effects of magnetic forces, can be expressed as follows:

$$\vec{f_{b_j}} = q_c \vec{E_j} \tag{3}$$

where $q_c$ is the plasma concentration in terms of $C/m^3$ and $\vec{E}$ represents the electric field vector.

The advanced electrostatic model [37] is one of the most physically comprehensive approximations for simulating the effect of a plasma actuator on fluid flow. The electrostatic model has been developed by assuming sufficient time for the continuous generation of plasma jets and quasi-steady-state behavior of the entire system [32]. Since the gas particles in the plasma generation process are weakly ionized, it is assumed that the electric potential consists of two components [32 and 37]: an electric potential due to the electric field and an electric potential due to charge density. Furthermore, due to the small Debye length and the thinness of the generated plasma layer, it is assumed that the plasma concentration is more influenced by the potential arising from charged particles on the walls and receives less influence from the external electric field. Using the principle of superposition, two main equations (4) and (5) are presented for the model [37]. Thus, by applying the mentioned assumptions in Maxwell's equations, two equations for the distribution of the electric potential field and charge density distribution are obtained:

$$\frac{\partial E_j}{\partial x_j} = \frac{\partial}{\partial x_j}\left(\varepsilon_r \frac{\partial \phi}{\partial x_j}\right) = 0 \tag{4}$$

$$\frac{\partial}{\partial x_j}\left(\varepsilon_r \frac{\partial q_c}{\partial x_j}\right) = \frac{q_c}{\lambda_d^2} \tag{5}$$

where, $\phi$ represents the electric potential, $\lambda_d$ is the Debye length, $\varepsilon_r$ is the relative permittivity, with a value of 1 for air and the specific value of 2.7 for the dielectric material used in this simulation, which is Kapton.

The electric potential equation (4) is solved in both the solid and fluid regions. The boundary conditions for solving the potential equation are as follows: On the outer boundary: $\partial \phi / \partial n\_i = 0$, on the outer electrode surface: $\phi = \phi(t)$, and on the inner electrode surface: $\phi = 0$. It is necessary to mention that $n_i$ in this equation is the unit normal vector to the surface, and $\phi(t) = \phi^{max} f(t)$ is defined to account for the alternating voltage variations applied.

The plasma density equation (5) is solved only in the fluid region, and the boundary conditions for solving this equation are as follows: On the outer boundary: $q_c = 0$, on a surface of the dielectric covering the inner electrode: $q_c = q_c^{max} G(x) f(t)$, and for the rest of the dielectric surfaces and the outer electrode: $\partial q_c / \partial n_i = 0$.

In the advanced model, to couple the two main equations (4) and (5), a boundary condition consistent with experimental results is used. The complete details of the advanced model can be found in the work of the authors [37].

$$\begin{aligned} 0 < x < 17\% \quad & q_c(x) = q_c^{max}\left(\frac{\phi_{max}^{local} - \phi}{\phi_{max}^{local} - \phi_{17\%}}\right)^{2.0} \\ 17\% < x < 100\% \quad & q_c(x) = q_c^{max}\left(\frac{\phi - \phi_{min}^{local}}{\phi_{17\%} - \phi_{min}^{local}}\right)^{0.6} \end{aligned} \tag{5}$$

Here, $\phi_{max}^{local}$ is the maximum electric potential over the charge surface and $\phi_{min}^{local}$ is the minimum electric potential over the charged surface, and $\phi_{17\%}$ is the electric potential at $x/l = 0.17$ from the leading edge of the exposed electrode with a length of l.

To solve the equations of the electric potential field and the plasma concentration distribution, it is preferable to nondimensionalize their equations and boundary conditions. This way, after solving the electric potential and charge concentration distributions for any voltage amplitude, can be obtained. The nondimensionalization process is evident in equations (7) and (8). By applying this nondimensionalization, the two governing equations of the electrostatic model are transformed into equations (9) and (10):

$$\phi^* = \phi/\phi(t) \tag{6}$$

$$q_c^* = q_c / q_c^{max} f(t) \tag{7}$$

$$\frac{\partial E_j^*}{\partial x_j^*} = \frac{\partial}{\partial x_j^*}(\varepsilon_r \frac{\partial \phi^*}{\partial x_j^*}) = 0 \tag{8}$$

$$\frac{\partial}{\partial x_j^*}\left(\varepsilon_r \frac{\partial q_c^*}{\partial x_j^*}\right) = \frac{q_c^*}{\lambda_d^2} \tag{9}$$

*3.3. Computational Procedure*

Simulation of flow over a cylinder is performed in an unsteady manner due to its inherent nature at a Reynolds number of 20,000. The unsteady simulations have been continued in all models until reaching a steady periodic state. It's worth mentioning that, considering the Reynolds number of the flow over the cylinder, the separation can be either laminar or turbulent. For a circular cylinder within Reynolds numbers ranging from 1,000 to 20,000, the flow is subcritical, signifying that the separation is fully laminar. However, at a Reynolds number of 20,000, although the separation is mostly laminar, the transition from laminar to turbulent is characterized by intermittent turbulent spots in the lower region of the cylinder. To observe the vortices and the boundary layer separation, the Turbulent Transition SST turbulence model is utilized. The working fluid is air under standard conditions, and its properties are used accordingly. In Table 1, comprehensive specifications of the flow solver for the desired numerical solution are presented.

**Table 1. Fluid flow solver methods**

| Simulation Part | Used Method |
| --- | --- |
| Navier-Stokes Type | Pressure-based |
| Pressure-Velocity Coupling | Coupled algorithm |
| Spatial Discretization | Second-order upwind |
| Temporal Discretization | Second-order |
| Gradient Calculations | Least squares cell-based |

*3.4. Computational Domain Geometries and Plasma Actuator*

The geometric dimensions of the plasma actuator installed on the cylinders with diameter of D are reported as follows: The electrode thickness is 0.01D, the dielectric thickness is 0.05D made of Kapton material with a relative permittivity of 2.7, the upper electrode length is 0.049D, and the lower electrode length is 0.089D. Two actuators are symmetrically installed on the cylinder with respect to the free-stream direction, as shown in Figure 1.

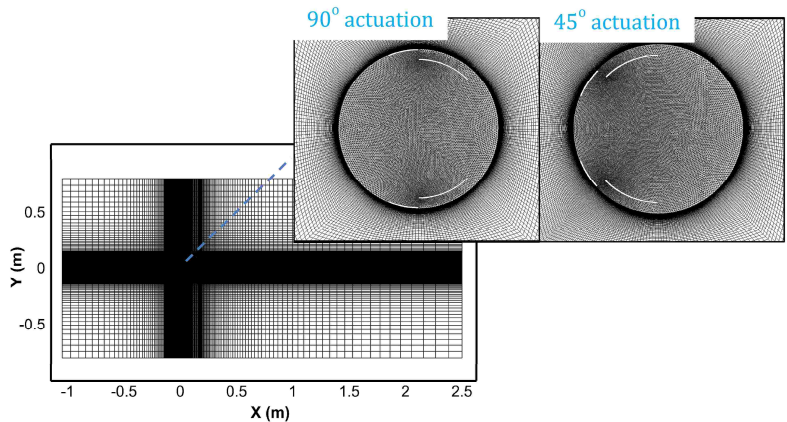

**Figure 1. Computational domain, structured generated mesh outside and inside of the cylinder**

As depicted in Figure 1, the numerical solution domain is a rectangle with dimensions 10 times the cylinder diameter above it, including the inlet boundary condition for the flow, and 25 times the cylinder diameter below it with the outlet boundary condition. The generated computational mesh outside the cylinders follows a structured grid with higher density near the walls, while the mesh within the cylinders is of a triangular type with increased density near the electrodes. The velocity-inlet boundary condition is employed for the flow inlet, and the outlet-vent boundary condition is used for the outlet of the computational domain. The upper and lower domain boundaries are set to periodic boundary conditions. The cylinder surface is treated as a smooth wall with a no-slip condition.

## 4. Computational Solver Validation

### 4.1. Fluid Flow Solver

According to numerical results presented in prior research [41], for the flow passing over a circular cylinder at Reynolds number of 20,000, the drag coefficient is 1.2 and the Strouhal number is 0.21. In the present study, for validation purposes, three coarse computational grids, medium, and fine resolutions have initially been generated. The outcomes of these three models are provided in Table 2. As evident in Table 2, considering the consistent variations of the Strouhal number for the fine grid, the generated grid with 52,410 cells and an acceptable average $Y^+$ value is chosen as the designated grid for the study. The time step size for this investigation is set at 0.001s. By comparing the outcomes of the current study with the experimental results [41], a discrepancy of 0.011 in the Strouhal number is observed, accounting for a 5.2% error. The error in calculating the mean drag coefficient is nearly negligible. For the purpose of numerical validation, the distribution of the time-averaged pressure coefficient on the cylinder is compared with the experimental results [41]. This comparison is depicted in Figure 2.

**Table 2. Grid study for computational simulation**

| Cell No | $C_d$ | Strouhal Number | Mean $Y^+$ |
|---|---|---|---|
| 35,246 | 1.38 | 0.190 | 1.80 |
| 52,410 | 1.18 | 0.221 | 1.20 |
| 81,156 | 1.16 | 0.225 | 1.14 |

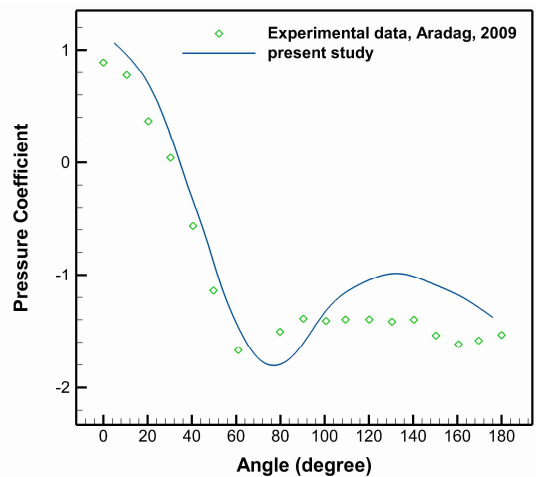

Figure 2. Comparison of the time-averaged pressure distribution with experimental results [41] (clockwise around the cylinder starting from the stationary point)

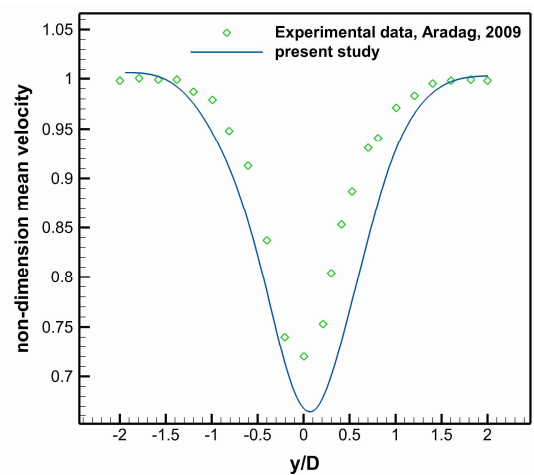

Figure 3. Comparison of the time-averaged velocity profile with experimental results [41] (vertical cross-section downstream from the cylinder at a distance of three cylinder diameters)

Starting from the stationary point and rotating clockwise around the cylinder, the distribution of the time-averaged pressure coefficient is plotted in Figure 2. This temporal average is calculated over a range of 20 cycle periods, spanning cycle periods 80 to 100. Additionally, the dimensionless time-averaged velocity distribution at a cross-section beneath the cylinder is compared with experimental results within the same cycle period range, as shown in Figure 3. This mean profile is measured at a location three diameters downstream from the cylinder. The validation results indicate that the numerical solution accuracy is suitable.

*4.2. Electrostatics Solver*

For the purpose of validating the modeling of the plasma actuator's effect using the electrostatic model, the flow of stationary air over a flat plate is simulated. The plate is excited, generating a wall jet through the plasma actuator. The jet velocity is 1m/s, as per experimental results by Jacob [42]. The applied voltage is 5 kV. The computational domain, as observed in Figure 4, is a square with a side length of 0.1m, and the plasma actuator is positioned at its center. The electrode length is 10mm, with a thickness of 0.102mm, and the dielectric thickness is 0.127mm. A horizontal gap of 0.05mm exists between the two electrodes.

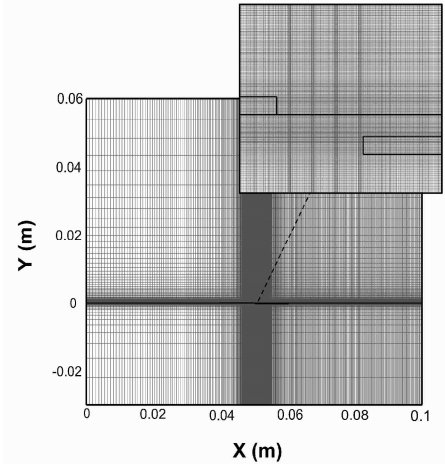 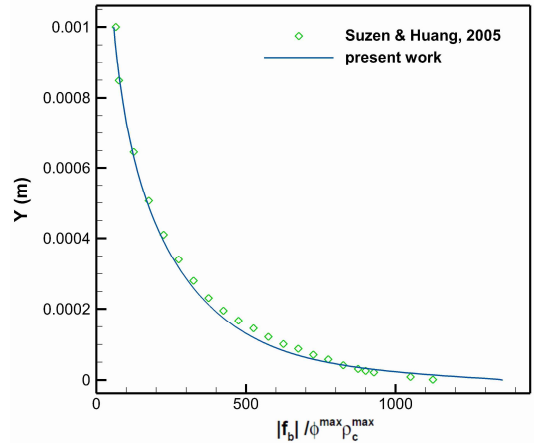

**Figure 4.** Computational domain and structured grid for the flat plate case

**Figure 5.** Comparison of electrostatic modeling results with the results of Suzen and Huang [32]

The SIMPLE algorithm with second-order accuracy is employed for solving, and the flow is assumed to be steady. Figure 5 presents a comparison of dimensionless body force results obtained from the current solution with the numerical solutions of Sozen and Huang [32] in a vertical section. The comparative results between the current numerical solution and other numerical solutions are reported in Table 3. As evident, the current solution exhibits reasonable accuracy for the modeling in comparison to the conducted numerical solutions.

**Table 3.** Comparison of numerical results for the induced flow over a flat plate with other numerical simulations [32 and 43]

| Properties | Present study | Suzen & Huang (2005) [32] | Bouchmal (2011) [43] |
|---|---|---|---|
| $F_b/\phi^m \rho^m$ | 1320 | 1250 | 1440 |
| $V_{jet}$ (m/s) | 0.938 | ≈1 | 0.934 |

## 5. Results and Discussion

*5.1. Single Cylinder Wake Control*

A two-dimensional flow passing over a circular cylinder at a Reynolds number of 20,000 is simulated. The applied voltage to the actuators is 10 kV, and the applied frequency is 6 Hz. Figure 6 illustrates the streamlines for three examined cases. Figure 7 further presents the profiles of horizontal velocity component. These profiles are depicted in a vertical section located 10 diameters downstream from the cylinder's lower end.

As seen in Figure 6, in case (A), the actuator is inactive, and velocity oscillations continue behind the cylinder over a substantial distance downstream, forming unstable vortices. These unstable vortices periodically detach from the cylinder and descend downstream in an alternating manner, generating a wave-like pattern in the flow lines downstream from the cylinder's lower end. As depicted in Figure 7, the velocity reaches a maximum reduction of 76% compared to the free-stream velocity.

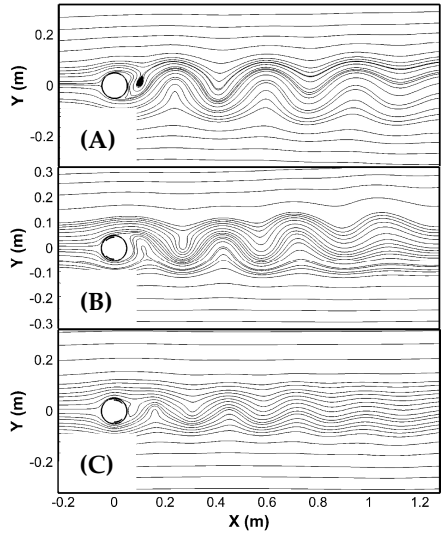

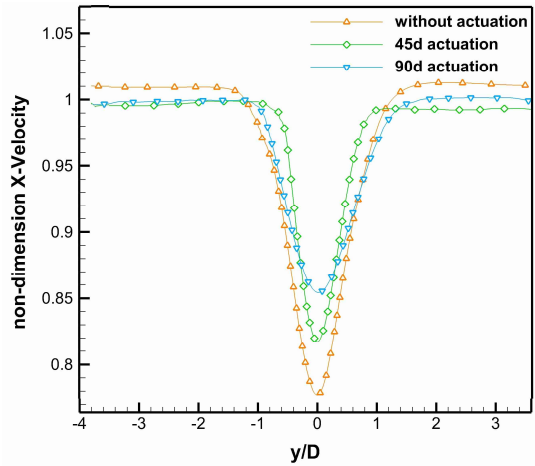

**Figure 6.** Comparison of streamlines around the single cylinder with and without the plasma actuator: (A) without the actuator, (B) at 45 degrees, (C) at 90 degrees

**Figure 7.** Comparison of downstream velocity profiles, at a distance of ten times the single cylinder diameter, with and without the plasma actuator

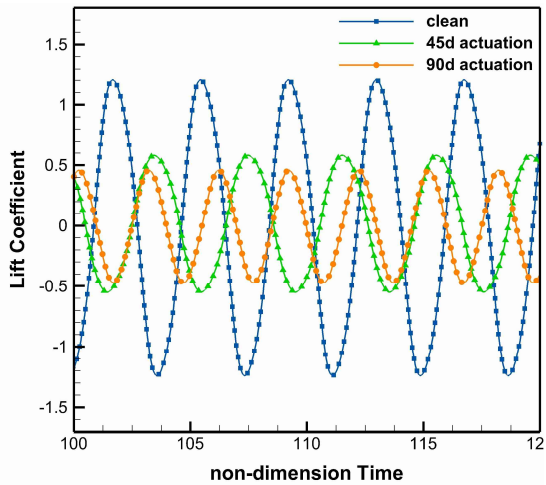

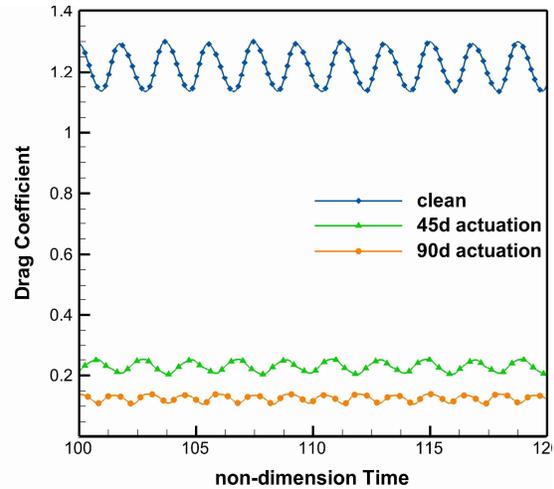

**Figure 8.** Comparison of the lift coefficient around the single cylinder with and without the plasma actuator

**Figure 9.** Comparison of the drag coefficient around the single cylinder with and without the plasma actuator

In case (B), by employing the actuator at a 45-degree angle and injecting momentum into the boundary layer at this region, it partially aids in preventing the formation of unstable vortices. Furthermore, a reduction in velocity fluctuations downstream can also be observed. However, due to the actuator's placement not being close to the precise separation point of the boundary layer, the effect of this actuator with the applied voltage of 10 kV is limited, and it might require a higher voltage to enhance the flow downstream. Figure 7 also shows a 6% improvement in the maximum velocity reduction behind the cylinder.

In case (C), the actuator is installed at a 90-degree angle. Placing the actuator at an angle that is relatively close to the separation point significantly contributes to stabilizing the flow behind the cylinder. As observed, velocity fluctuations reach a more uniform state much quicker than before. This effect can also be seen in Figure 7. A 10% improvement in the maximum velocity reduction behind the cylinder validates this outcome.

Figures 8 and 9 illustrate changes in the lift coefficient ($C_L$) and the drag coefficient ($C_D$) on the cylinder for three examined states. As depicted in Figure 8, the use of the plasma actuator has reduced the range of lift force variations. This reduction is more pronounced when placing the actuator at a 90-degree angle compared to placing it at a 45-degree angle. In Figure 9, it can also be observed that the changes in the drag coefficient have been influenced by the placement of the plasma actuator. The decrease in the drag coefficient compared to the case without using the plasma actuator is evident in this figure. By employing the plasma actuator at a 90-degree angle, due to the proximity of its placement to the starting position of flow separation, we witness a lesser wake production.

*5.2. Tandem Cylinder Wake Control*

In the simulation of the second part, two cylinders of equal diameters are placed at a center-to-center distance of four times their diameter. The Reynolds number of the free-stream flow in this setup is 20,000. In this section, the effect of applying an actuator with varying strengths on the upstream cylinder is investigated regarding the flow quality behind the downstream cylinder. The actuators are symmetrically positioned relative to the free-stream direction and are installed at a 90-degree angle on the cylinder.

Figure 10 presents the streamlines for the four examined models in this section. One of these cases, (A) represents the flow passing through the tandem cylinders without using an actuator, and the other three models involve the use of two plasma actuators installed at a 90-degree angle on the upstream cylinder with different voltages: 10 (B), 13 (C), and 16 (D) kV. Additionally, in Figure 11, profiles of the horizontal velocity component are provided in a vertical section located 10 times the cylinder diameter downstream from the lower cylinder.

In case (A), the actuator is inactive. The unstable flow passing through the upper cylinder with its own oscillations and instabilities collides with the lower cylinder, thus creating a substantial velocity reduction behind the lower cylinder. As observed in Figure 10, unstable vortices continue to exist behind the lower cylinder up to a distance of 5 times the cylinder diameter. Consequently, a significant velocity reduction is established behind the lower cylinder. Furthermore, in Figure 11, it can be observed that at a distance of 10 times the cylinder diameter downstream from the lower cylinder, the maximum velocity reduction has reached 53% of the free-stream velocity.

In case (B), the actuators installed above and below the upper cylinder, at a 90-degree angle, with a voltage of 10 kV, are controlling the flow passing through the upper cylinder. The reason for positioning the actuators at a 90-degree angle in these models is their placement at the closest separation points of the boundary layer, as indicated by the results from the previous section. The use of actuators in this configuration leads to a flow encountering the lower cylinder with reduced velocity oscillations. This results in a reduction in velocity reduction and a decrease in the length of unstable vortices behind the lower cylinder. As shown in Figure 11, the velocity reduction has been significantly improved, reaching a 20% reduction in the maximum value.

In cases (C) and (D), the voltages applied to the actuators are 13 and 16 kV, respectively. As observed, the flow passing through the upper cylinder has become completely uniform with the application of a 16-kV voltage. However, as shown in Figure 10, there aren't substantial changes compared to model (C), where a 13-kV voltage was applied. This observation is consistent in Figure 11 as well. A 3% improvement is visible for the 13-kV voltage application, and a 6% improvement is observed for the 16-kV voltage application relative to the state with a 10-kV voltage application, as depicted in this figure.

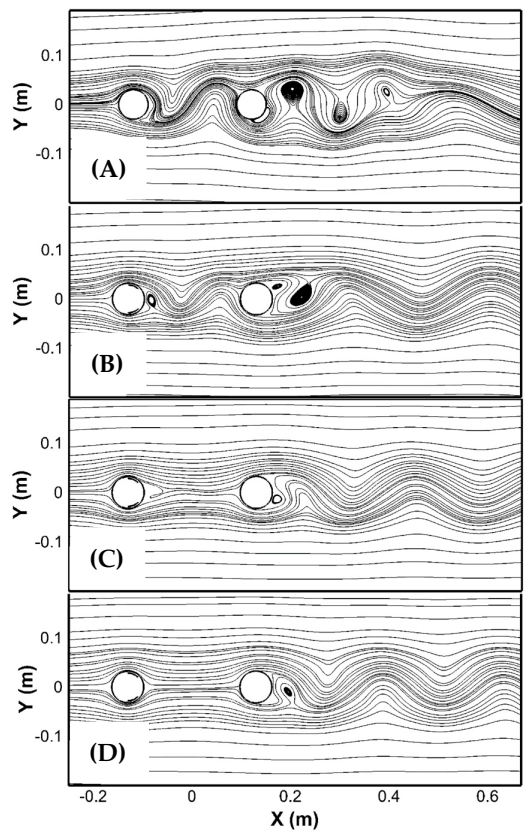
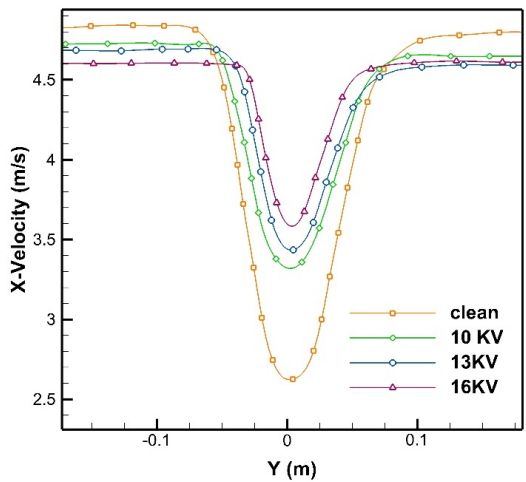

Figure 10. Comparison of streamlines around the tandem cylinders with the use of a plasma actuator:
(A) Without the actuator,
(B) Actuator with 10-kV voltage,
(C) Actuator with 13-kV voltage,
(D) Actuator with 16-kV voltage

Figure 11. Comparison of velocity profiles downstream of the tandem cylinders with and without the actuator at different voltage levels

## 6. Conclusion

The utilization of DBD plasma actuators has garnered significant attention due to their low power consumption. A notable point regarding the use of these actuators is that their effectiveness strongly depends on their appropriate placement and suitable power. In the conducted simulation, the primary focus was on employing an advanced electrostatic numerical model for simulating the flow passage and its control. The aim was to investigate the functionality of this actuator accurately without expensive laboratory tests. The use of a 10-kV, 6-Hz operation on a single cylinder in two different positions revealed that injecting momentum near the 90-degree angle yields the best results at Reynolds number 20,000. Interestingly, the actuator placed at a 45-degree angle with the same voltage could not significantly impact flow quality control. Moreover, noteworthy improvements were observed in the fluctuations of the lift and drag coefficients. In the second part of this study, the impact of the flow quality from the first cylinder on a flow passing through a tandem cylinder was examined. With an increase in the applied voltage, an enhanced flow quality was observed downstream of the second cylinder.